%% file: Main.tex
\documentclass[conference]{IEEEtran}
\IEEEoverridecommandlockouts
\usepackage{cite}
\usepackage{amsmath,amssymb,amsfonts}
\usepackage{graphicx}
\usepackage{textcomp}
\usepackage{xcolor}
\usepackage{algorithm}
\usepackage{subcaption}
\usepackage{algpseudocode}
\usepackage{hyperref}
\usepackage{newtxmath} % for Times Roman clone math font
\usepackage[scr=rsfso]{mathalfa} % "oblique" version of mathrsfs
\usepackage[acronym]{glossaries}

\newacronym{design}{NirvaWave}{mmWave and sub-THz Near Field Wave Propagation Simulator for 6G and Beyond}
\def\BibTeX{{\rm B\kern-.05em{\sc i\kern-.025em b}\kern-.08em
    T\kern-.1667em\lower.7ex\hbox{E}\kern-.125emX}}
\begin{document}

\title{NirvaWave: An Accurate and Efficient Near Field Wave Propagation Simulator for 6G and Beyond\\
\thanks{This work was supported by NSF under Grant CNS-2145240.}
}

\author{\IEEEauthorblockN{Vahid Yazdnian}
\IEEEauthorblockA{\textit{Electrical Engineering Department} \\
\textit{Princeton University}\\
Princeton, USA \\
vahidyazdnian@princeton.edu}
\and
\IEEEauthorblockN{Atsutse Kludze}
\IEEEauthorblockA{\textit{Electrical Engineering Department} \\
\textit{Princeton University}\\
Princeton, USA \\
kludze@princeton.edu}
\and
\IEEEauthorblockN{Yasaman Ghasempour}
\IEEEauthorblockA{\textit{Electrical Engineering Department} \\
\textit{Princeton University}\\
Princeton, USA \\
ghasempour@princeton.edu}

}
\maketitle
\begin{abstract}
The extended near-field range in future mm-Wave and sub-THz wireless networks demands a precise and efficient near-field channel simulator for understanding and optimizing wireless communications in this less-explored regime. This paper presents \acrshort{design}, a novel near-field channel simulator, built on scalar diffraction theory and Fourier principles, to provide precise wave propagation response in complex wireless mediums under custom user-defined transmitted EM signals. \acrshort{design} offers an interface for investigating novel near-field wavefronts, e.g., Airy beams, Bessel beams, and the interaction of mmWave and sub-THz signals with obstructions, reflectors, and scatterers. The simulation run-time in \acrshort{design} is orders of magnitude lower than its EM software counterparts that directly solve Maxwell Equations. Hence, \acrshort{design} enables a user-friendly interface for large-scale channel simulations required for developing new model-driven and data-driven techniques. We evaluated the performance of \acrshort{design} through direct comparison with EM simulation software. Finally, we have open-sourced the core codebase of \acrshort{design} in our GitHub repository.\footnote{The GitHub repository of \acrshort{design} can be found at \href{https://github.com/vahidyazdnian1378/NirvaWave}{https://github.com/vahidyazdnian1378/NirvaWave}.} 

 % This paper introduces \acrshort{design}, a novel near-field channel simulator based on scalar diffraction theory, designed to explore EM wave propagation in diverse environments. \acrshort{design} accurately models interactions with blockages, reflectors, and Reconfigurable Intelligent Surfaces (RIS), supporting multi-Tx antenna arrays with arbitrary phase configurations. It features a user-friendly interface and open-source code, validated against EM simulators, to facilitate AI-driven solutions. The core codebase is accessible for further development.

\end{abstract}
\vspace{+1mm}
\begin{IEEEkeywords}
Near-Field, Simulator, 6G, sub-THz, mmWave
\end{IEEEkeywords}

\input{introduction}
\input{design}

\input{evaluation}

 \input{conclusion}
 
\bibliographystyle{IEEEtran}
\bibliography{references}

\end{document}

%% file: introduction.tex
\vspace{-1mm}
\section{Introduction}
Millimeter-wave (mmWave) and Sub-Terahertz (sub-THz) communications are emerging as promising solutions to the spectrum scarcity faced by traditional WiFi and cellular networks~\cite{tataria20216g}. The abundant bandwidth in these frequency bands offers the potential to achieve multi-Gbps data rates essential for immersive extended/virtual reality, high-resolution streaming, intelligent autonomous systems, and ultra-low-latency backhauling\cite{tataria20216g}. However, the severe propagation loss inherent to this spectral regime necessitates the use of larger apertures for transmitters and receivers, like fixed directional antennas, phased arrays, and reconfigurable intelligent surfaces (RIS), to create highly directional beams to compensate for path loss.

The large array aperture combined with high carrier frequencies in the mmWave and sub-THz bands yields the extension of the near-field region to several meters causing a paradigm shift, i.e., positioning many receivers in the near-field region of the transmitter antennas. 
Unfortunately, the existing channel simulators are either built on the assumption of planar-wave far-field transmissions or are prohibitively time-consuming for realistic simulations. 

Specifically, EM solvers such as Altair Feko~\cite{feko} and CST Studio Suite offer accurate EM wave propagation of complex antenna apertures at any frequency and can capture the wave interaction with arbitrarily designed reflectors/obstruction. This is achieved by solving Maxwell's equations using various techniques, including the Method of Moments (MoM), Finite Difference Time Domain (FDTD), and Finite Element Method (FEM)~\cite{balanis2016}. Although these solvers provide accurate solutions, they lack scalability, as their run-time grows exponentially with the aperture size and the complexity of the wireless medium. Indeed, even simulating simple environments may take up to several hours. Less demanding channel simulators either operate based on statistical channel models (e.g., 3GPP~\cite{samimi2016local} and NYUSIM~\cite{sun2017novel}) or exploit ray tracing methods (e.g., Sionaa developed by NVIDIA~\cite{10465179}). Unfortunately, these schemes do not take into account the spherical wavefront radiated from every element in the transmitting array, and hence are inaccurate in the near-field region. Therefore, developing a precise and efficient near-field channel simulator is crucial for advancing our understanding of mmWave and sub-THz communications in the next-generation wireless networks.

This paper presents \acrshort{design}, an open-source channel simulator developed for accurate and efficient modeling of electromagnetic wave propagation in near-field wireless settings. \acrshort{design} leverages the scalar diffraction theory and the Rayleigh-Sommerfeld integral to ensure a high level of accuracy by adhering completely to the fundamental physics of EM wave propagation. To achieve computational efficiency, \acrshort{design}'s core algorithm is built based on the Angular Spectrum Method that exploits Fourier properties.
%to efficiently calculate EM wave propagation in complex mediums.
\acrshort{design} supports free-space and inhomogeneous near-field mediums with blockers and reflectors. The complex electric field at the transmitter (including phase and amplitude per antenna) can be defined by the user. We have implemented emerging near-field wavefronts with unique characteristics as default options. This includes conventional beam shaping schemes such as far-field Gaussian beams and near-field focused beams as well new wavefront like Airy beams~\cite{latychevskaia2016creating} with curved trajectories and Bessel beams~\cite{singh2022bessel} with the unique property of self-healing, i.e., the wavefront is reconstructed after interacting with obstruction. We emphasize that today's wireless channel simulators do not allow for the implementation of these exotic wavefronts. Additionally, the simulator supports RIS implementation and can capture scattering off of rough surfaces with wavelength-level height perturbation. Finally, users can define realistic wireless mediums with multiple TXs, RXs, and obstructions and find the received signal heatmap in such complex settings.

We rigorously benchmark \acrshort{design} EM wave propagation simulations against highly accurate yet computationally demanding EM simulators. Our extensive simulation results demonstrate strong alignment with the ground truth across various environments while reducing the simulation run-time by orders of magnitude. 
We believe \acrshort{design} provides a user-friendly interface for studying mmWave and sub-THz near-field channels and will pave the way for developing model-driven and data-driven techniques to address the challenges of wireless communications in these high-frequency regimes.

%% file: design.tex
\section{Near-Field Channel Modeling and \\ Simulator Design}
 In this section, we explain the physical principles incorporated into \acrshort{design} making it capable of accurately and efficiently modeling near-field EM wave propagation. Specifically, we will first explain the Rayleigh-Sommerfeld integral theory that serves as the foundational principle for modeling the radiation of spherical waves in free-space. Then, we extend our modeling to inhomogeneous environments with the presence of blockers and reflectors. We show how small perturbations in the surface of the reflector can cause diffuse scattering and its impact on reflection characteristics. We will also explain the principles of RIS implementation in near-field channels. Finally, we outline the design flow of the simulator and describe \acrshort{design}'s core algorithm.

\input{rayleigh}

\input{blockage}

\input{reflection}

\input{rough}

\input{RIS}
\input{coordinate_algorithm}

\input{GUI}

%% file: rayleigh.tex
\subsection{Primer on Rayleigh-Sommerfeld Integral Theory}
\begin{figure}[t]
    \centering
    \includegraphics[width=0.3 \textwidth]{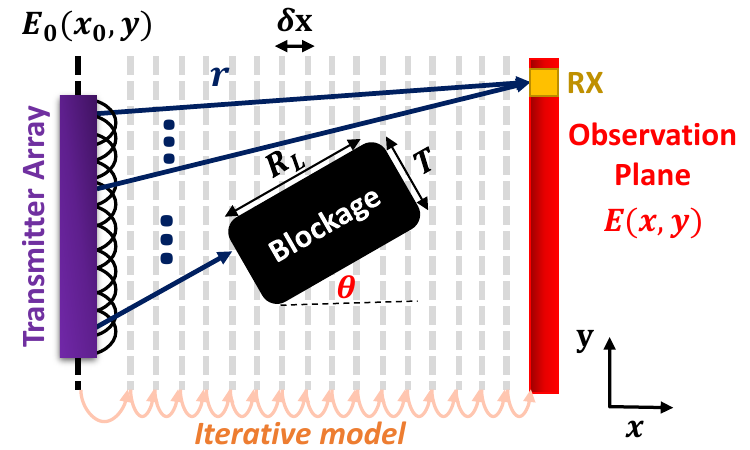}
    \vspace{-1mm}
    \caption{Demonstration of the iterative Rayleigh-Sommerfeld integral theory that captures the evolution of an arbitrary electric field distribution $E_{0}(x_0,y)$ in space in the presence of blockers and reflectors.}
    \label{fig:RS}
    \vspace{-5mm}
\end{figure}

Assume we have a transmitter array at $x=x_0$, as shown in Fig.~\ref{fig:RS}. The Rayleigh-Sommerfeld theory suggests that we can consider this aperture as an infinite number of point sources each emitting spherical waves. Further,  to find the electric field in an arbitrary point (x,y,z), we have to add the contributions from all these point sources together. Since in this theory, there is no assumption about the initial E-field distribution,  it allows us to implement any phase/amplitude profile on TX antenna arrays and find the corresponding electric field at arbitrary RX locations or observation planes. The general form of the Rayleigh-Sommerfeld integral can be derived from Maxwell's equations by use of Green's Theorem as follows~\cite{delen1998free}:
\begin{equation}\label{SommerfeldIntegral}
\begin{aligned}
        &E(x', y', z'|E_0) = \\
       &\frac{1}{2\pi} \iint_{A} E_0(x_0, y, z) \left( \frac{\Delta x \, e^{-ik r}}{r^2} \left(i k + \frac{1}{r}\right) \right) \, dydz,
\end{aligned}
\end{equation}
%where, $E(x_0, y, z)$ represents the electric field at $x_0$ which is considered to be the initial E-field distribution.
where, $E(x_0, y, z)$ represents the initial E-field distribution at $x_0$. This integral theory would find the complex E-field at any arbitrary point $(x', y', z')$. $k$ represents the wave number, $\Delta x = x'-x_0$, and $r = \sqrt{ (x' - x_0)^2 + (y' - y)^2+(z' - z)^2}$ is the distance from each source point to the desired location. 

Although the Rayleigh-Sommerfeld integral provides a straightforward method to find the EM wave evolution as it propagates between TX and RX, it requires computing a discrete integral at each point in space, which is significantly time-consuming. The Angular Spectrum Method (ASM) simplifies the computation of this integral by transforming the equation into the frequency domain using Fourier principles, leveraging the fact that Eq.~(\ref{SommerfeldIntegral}) can be written in the form of convolution. Particularly, using ASM, we can write:
%even in free-space, particularly when the spatial resolution is high (i.e., small values of $\delta x$ in Fig.~\ref{fig:RS}).
\begin{equation}\label{ASM}
\begin{aligned}
    \mathcal{F}\{ E(&x',y',z'|E_0)\} = \\
    &\mathcal{F}\{ E(x_0,y,z)\}  \times \mathcal{F}\left\{ \frac{1}{2\pi} \frac{\partial}{\partial x'} \left( \frac{e^{-ik r}}{r} \right) \right\}
\end{aligned}
\end{equation}
This way we can interpret the free space propagation as a linear system in which $\mathcal{F} \left\{ \frac{1}{2\pi} \frac{\partial}{\partial x'} \left( \frac{e^{-ik r}}{r} \right) \right\}$ can be viewed as a transfer function that captures the EM disturbance of a point source transmission. In other words, we can write:
\begin{equation}\label{transferfunction}
\begin{aligned}
        & H(f_y,f_z) = \mathcal{F}\left\{ \frac{1}{2\pi} \frac{\partial}{\partial x'} \left( \frac{e^{-ik r}}{r} \right) \right\}= \\
        & exp(-i2\pi\frac{(x'-x)}{\lambda}\sqrt{1-\lambda^2\;( f_y^2+f_z^2)}),
\end{aligned}
\end{equation}
where $\lambda$ is the wavelength, $f_y$ and $f_z$ represent spatial frequencies. We highlight that the transfer function $H(f_y,f_z)$ is solely a function of the geometry of the environment and not the transmitted electric field. Finally, $ E(x',y',z'|E_0)$ can be calculated by taking a Fourier inverse as follows:
\begin{equation}\label{finalRS}
E(x',y',z'|E_0)= \mathcal{F}^{-1}\left\{\mathcal{F}\{ E(x_0,y,z)\}\times H(f_y,f_z)\right\}. 
\end{equation}
In \acrshort{design}, we use the simplified 2D version of Eq.~(\ref{SommerfeldIntegral})-(\ref{finalRS}). Indeed, Fourier transforms are much more computationally efficient and help reduce the complexity and rum-time of simulations in \acrshort{design}. However, as mentioned earlier, such calculation only applies to free-space communication and other techniques needed to extend it to a wireless medium that involves blockers and reflectors. 

%$f_y=\frac{sin(\theta_y)}{\lambda}$, $f_z=\frac{sin(\theta_z)}{\lambda}$

%% file: blockage.tex
\subsection{Near-Field Blockage Modeling and Simulation}
The Rayleigh-Sommerfeld integral theory originally describes EM wave propagation in free space. However, to account for diffraction due to environmental blockages, we must model the relevant boundary conditions.  
In free space, using the angular spectrum method, we can compute the electric field at each location $x=x'$ either based on the initial E-field at $x=x_0$ in one shot or based on the previously calculated E-field at $x=x'- \delta x$ using an iterative approach.
However, when there are blockages in the environment, we must rely on an iterative scheme to account for any disturbances and discontinuities in the E-field caused by the presence of blockers. 

As depicted in Fig. \ref{fig:RS}, $\delta x$ represents the discrete step size for iterative calculations, and $R_L$ and $T$ denote blocker length and thickness respectively. In \acrshort{design}, we use $BL(x,y)$ to characterize arbitrary-shaped blockers. Specifically, if $BL(x,y)=1$, the signal is unattenuated at point $(x,y)$, indicating no blocker. Conversely, $BL(x,y)=\alpha$, where $0 \le \alpha<1$, captures the attenuation constant caused by a blocker at (x,y). 
Thus, the E-field distribution at the observation plane can be determined through an iterative process with the $k$th iteration for $x=x_0+k\delta x$ expressed as:
\begin{equation}\label{iterative}
    E(x,y) = BL(x,y)\times \mathcal{F}^{-1}\{ H(f_y)\times \mathcal{F}\{ E(x-\delta x,y)\}\}
\end{equation}
Using this approach, \acrshort{design} is able to account for the diffraction behavior of EM wave propagation in the presence of blockages in a time-efficient manner.

%% file: reflection.tex
\subsection{Near-Field Reflection Modeling and Simulation}\label{reflection}
For simplicity and without loss of generality, we will first explain how to characterize a single near-field reflector in the environment, followed by a discussion of the general case using a recursive algorithm. To this end, \acrshort{design} starts by calculating the E-field in the environment assuming there are no reflections. Then, we can find the electric field incident on the reflector's surface denoted as $E(x_{ref},y_{ref})$, where $(x_{ref},y_{ref})$ represents the points on the reflector. Building on top of the Huygens-Fresnel principle~\cite{depasse1995huygens}, we treat the reflector as another source of EM signals in the medium, determined using the previously calculated initial radiating E-field. In other words, similar to modeling the transmitter, we consider the reflector as an infinite number of point sources radiating spherical wavefronts. Therefore, the reflected signal can also be characterized by the Angular Spectrum Method. 
\begin{figure}[b]
    \centering
    \vspace{-5mm}
    \includegraphics[width=0.3\textwidth]{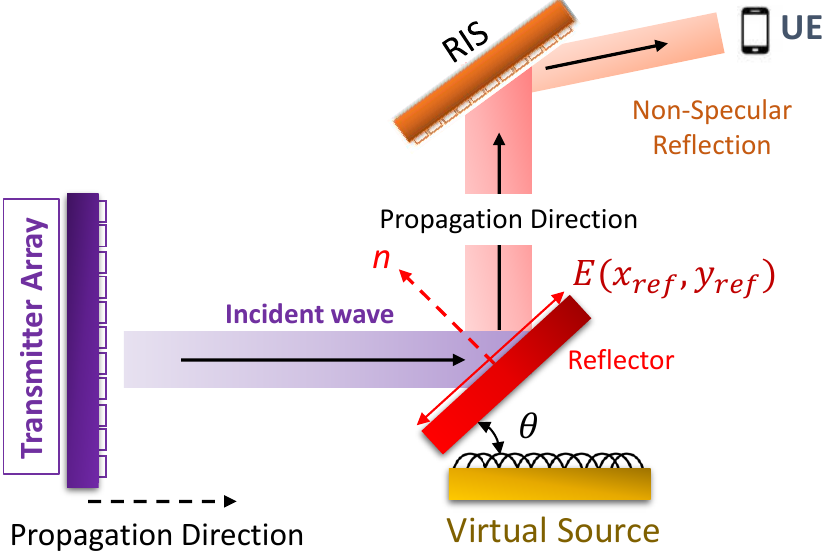}
     \vspace{-2mm}
    \caption{Demonstration of reflection and reconfigurable intelligent surfaces implementation in \acrshort{design}.}
    \label{fig:ref}
\end{figure}

It is important to note that ASM originally describes EM wave propagation when the direction of propagation is normal to the E-field source plane. However, when considering reflections, the propagation direction is no longer normal to the reflector plane. Accordingly, the original transfer function described in Eq.~(\ref{transferfunction}) must be modified. Indeed, the rotation angle between reflector and virtual source planes (denoted as $\theta$ in Fig.~\ref{fig:ref}) is purely a function of reflector orientation relative to TX.  Hence, the transfer function in Eq.~(\ref{transferfunction}) can be expressed based on full diffraction theory as~\cite{delen1998free}:
%The first step in modifying the transfer function is to determine the tilting angle between the virtual plane source of propagation, whose normal is aligned with the direction of the reflected EM wave. According to the reflection boundary condition\cite{}, each source of spherical waves at the transmitter (Tx) incident on the reflector plane can be treated as a secondary point source radiating spherical waves from the image point relative to the reflector plane. Thus, the virtual plane source can be found by reflecting the original source with respect to the reflector plane. We can write down the tilting angle as follows:
%$$\theta = \begin{cases}
%|\theta_R| & \text{if } |\theta_R| \leq \frac{\pi}{4}\\
%\frac{\pi}{2}-|\theta_R| & \text{if } |\theta_R| > \frac{\pi}{4}
%\end{cases}$$
%In this equation $\theta_R$ represent the reflector orientation and satisfy $-\pi/2<\theta_R<\pi/2$. Now we can modify the transfer function using the free-space beam propagation between arbitrarily oriented planes based on full diffraction theory \cite{} as follows:
\begin{equation}\label{transferfunction_modified}
\begin{aligned}
        H_{m}(f_y|\theta) = exp(-i2\pi\frac{(x'-x)}{\lambda cos(\theta)}\sqrt{1-(\lambda\;f_y \, cos(\theta))^2} \,)
\end{aligned}
\vspace{-1mm}
\end{equation}

Therefore, the EM wave propagation of the reflected wave can be derived using Eq.~(\ref{finalRS}) by considering the modified transfer function $H_m(f_y|\theta)$ and the virtual initial EM wave source $E_{vir}(x_0,y) = E(x_{ref},y_{ref})$. Hence, we can compute the reflected E-field profile in the reflector coordinate system (similar to Eq.~(\ref{iterative})) through an iterative process:
\begin{equation}\label{reflection_modeling}
\begin{aligned}
        E_{vir}(x,y) =  &BL_{ref}(x,y) \times \\
        &\mathcal{F}^{-1}\{ H_{m}(f_y|\theta)\times \mathcal{F}\{\Gamma_{r} \times E_{vir}(x-\delta x,y)\}\}
\end{aligned}
\end{equation}
where $BL_{ref}(x,y)$ denotes the blocker properties in the reflector coordinate system and $\Gamma_{r}$ is the reflection coefficient (between 0-1) that can be input by the users. Using this approach, we can model the reflected EM wave propagation in the coordinates of the corresponding reflector plane. Later in Sec.~\ref{algorithm}, we will explain how to find the total field propagation considering the contribution of the TX radiation and other potential virtual sources (i.e., reflectors/RISs).

%% file: rough.tex
\subsection{Diffuse Rough Scattering Implementation in \acrshort{design}}
In Sec.~\ref{reflection}, we explained how a specular reflection can be captured in \acrshort{design}. In practice, reflection at high frequencies also includes diffuse scattering components as the small perturbations on the surface of the reflection become comparable with the sub-mm wavelength of the impinging waves. Indeed, past work showed such scattering behavior has important implications for signal coverage and mobility resilience in sub-THz wireless networks~\cite{shen2023scattering}. Hence, here we explain how to model rough scattering surfaces in \acrshort{design}.

Surface perturbations are often modeled with a Gaussian distribution, with height at location $(x)$ as $H(x) \sim \mathcal{N}(0, h_{rms}^2)$, where $h_{rms}$ is the standard deviation of the surface height. Similarly, we can define correlation length $Lc$ as an indicator of horizontal roughness~\cite{fung1994microwave}. \acrshort{design} allows users input ($h_{rms}$, $L_c$) parameters that capture the statistical profile of random diffuse scattering to analyze the effects of that on communication systems. Specifically, a generated random height perturbation $H (x,y)$ based on user-defined $h_{rms}$ and $L_c$ values is translated to the corresponding additional phase variation on the suraface as: $\phi_{rough} (x,y; h_{rms}, L_c) =  2\pi \frac{H (x,y)}{\lambda}$.
% Specifically, a random surface is generated by the user-defined $h_{rms}$ and $L_c$ values, and then the height perturbation is translated to the corresponding additional phase shifts as: $\phi_{rough} (x,y; h_{rms}, L_c) =  2\pi \frac{H (x,y)}{\lambda}$, where $\phi_{rough} (x,y)$ characterizes the phase variation due to minute random surface height $H (x,y)$. 
Therefore, the electric field reflected from the surface of a rough object can be approximated as:
\begin{equation}
\label{eq:E_rough}
         E_{vir}(x_{0},y) = E(x_{ref},y_{ref}) \times e^{j2\pi \phi_{rough} (x,y; h_{rms}, L_c)} 
 \end{equation}
 Hence, the reflection off of a rough surface can be characterized similar to a smooth reflector, albeit by updating the electric field at the virtual source according to Eq.~(\ref{eq:E_rough}) above.

%% file: RIS.tex
\subsection{Implemenation of Reconfigurable Surfaces in \acrshort{design} }
Reconfigurable Intelligent Surfaces (RIS) are emerging as a promising technology for future wireless communication systems. RIS employs a large number of low-cost passive unit cells to intelligently and flexibly control electromagnetic wave properties—such as amplitude, phase, and polarization—thereby optimizing the propagation channel to establish the best possible transmission links. Therefore, implementing near-field channel modeling to analyze THz and sub-THz wireless systems in the presence of RIS would be highly important. To implement RIS, we need to change the phase and amplitude configuration of the calculated E-field on the RIS plane, $E(x_{RIS},y_{RIS})$, based on the user-defined RIS phase shift and amplitude for each element, denoted by $\phi_{RIS}$ and $A_{RIS}$. Specifically, we can write:
\begin{equation}
    E_{vir}(x_{0},y) = E(x_{RIS},y_{RIS}) \times A_{RIS} e^{j 2\pi \phi_{RIS}}
\end{equation}
Using this approach, \acrshort{design} is able to model EM wave reflection off of the RIS by modifying the reflecting E-field distribution from $E_{vir}(x_{0},y)$ to $E(x_{RIS},y_{RIS})$ in Eq.~(\ref{reflection_modeling}).

%% file: coordinate_algorithm.tex
%  \begin{algorithm}[t]
% \small
% \caption{\acrshort{design} Recursive Algorithm}\label{algorithm}
% \begin{algorithmic}
% \State \textbf{Input:} List of Objects $Obj$, Primary E-field Distribution $E_0$
% \State \textbf{Output:} Near-field electric field distribution $E_{out}$
% \Function{total\_propagation}{$Obj$, $E_0$}
%     \If{depth condition is true}
%         \State $E_{out} \gets$ primary\_propagation($Obj$, $E_0$)
%         \For{each reflector in $Obj$}
%             \State $Obj_{ref} \gets$ transform\_coordinates($Obj$, reflector)
%             \State $E_{vir}(x_{0},y) \gets$ find\_E\_on\_reflector($E_{out}$, reflector)
%             \If{$|E_{vir}(x_{0},y)|^2> \text{threshold power}$}
%                 \State $E_{vir} \gets$ total\_propagation($Obj^*$, $E_{vir}(x_{0},y)$)
%                 \State $E_{vir(trans)} \gets \text{E\_field\_original\_coordinate($E_{ref}$)}$
%                 \State $E_{out} \gets E_{vir(trans)} + E_{out}$
%             \EndIf
%         \EndFor
%     \EndIf
% \EndFunction
% \end{algorithmic}
% \end{algorithm}

\subsection{\acrshort{design} Recursive Algorithm} \label{algorithm}
Thus far, we have provided the fundamental physical principles to model the near-field EM wave propagation in complex inhomogeneous mediums. Here, we explain the core algorithm of \acrshort{design}. In general, the environment may include several blockers, reflectors, or RISs. Hence, \acrshort{design} finds the wave propagation through a recursive algorithm, where the function is called within itself to account for the reflections off of the consecutive reflectors/RIS planes. %\acrshort{design}'s core algorithm is summarized in algorithm \ref{algorithm}.
This recursive algorithm first solves the EM wave propagation based on the initial electric field on the TX array and finds incident E-field on both sides of all reflector/RIS planes. This is done using a list of objects in the field of view and their geometric features. Then, for each reflector/RIS in the environment, the object list is updated using coordinate transformation and the contribution of the reflectors (as virtual sources) is calculated by repeating the same procedure. The algorithm would continue calculating the consecutive reflections until the termination condition is satisfied. Ultimately, the transformed E-fields resulting from the reflections and the TX radiation are recursively summed to obtain the total electric field in the environment.
% 

% \begin{figure}[htbp]
%     \centering
%     \includegraphics[width=0.5\textwidth]{figures/coverage_map.pdf}
%     \caption{Examples of near-field coverage map in \acrshort{design} simulator in the presence of reflectors/blockages with Airy beam and beam steering TX antenna configurations.}
%     \label{fig:coverage_map}
% \end{figure}

%% file: GUI.tex
  \section{Graphical User Interface}
\begin{figure}[b]
    \centering
    \vspace{-4mm}
    \includegraphics[height=6.5cm, width=0.45\textwidth]{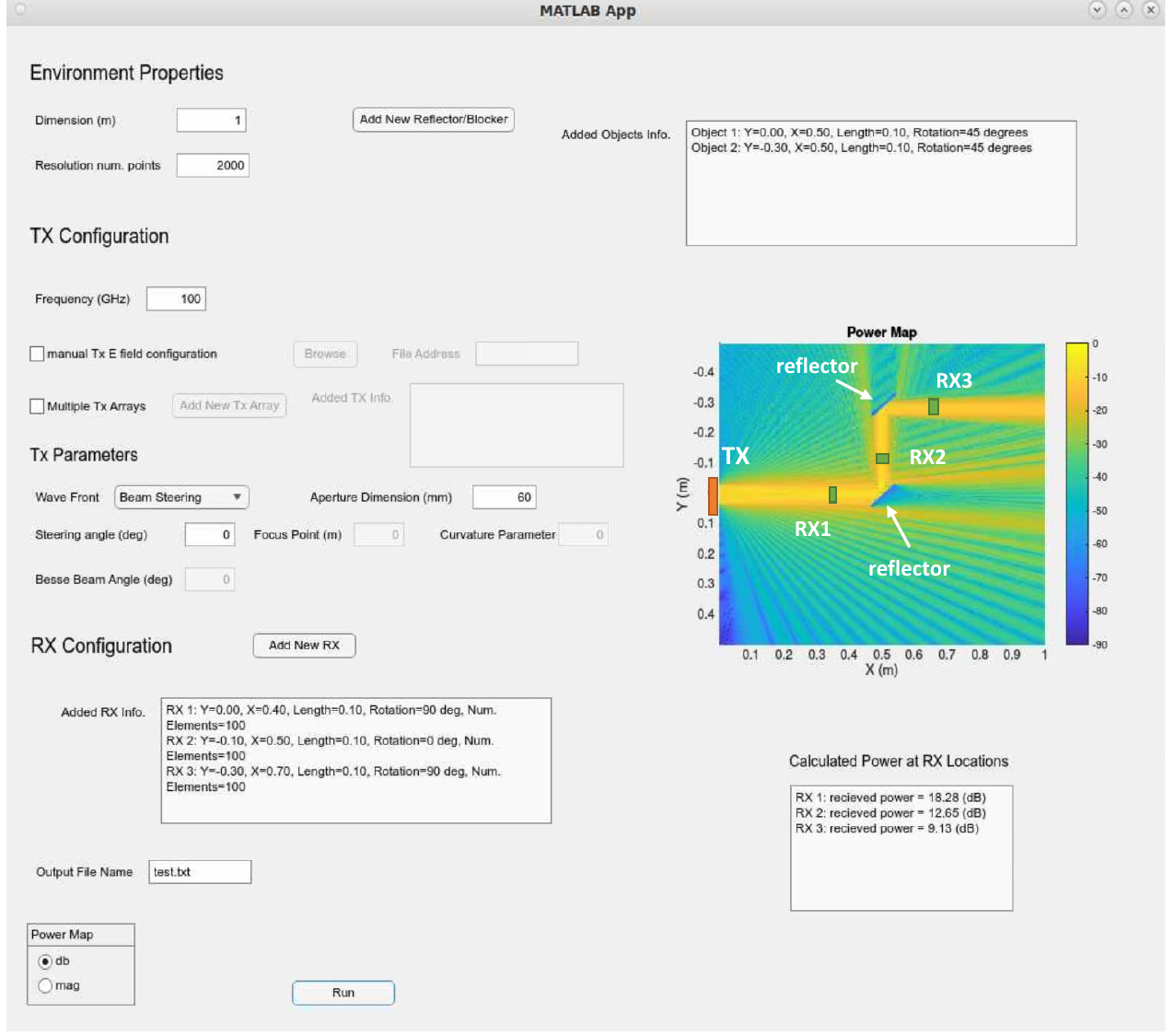}
      \vspace{-2mm}
    \caption{Graphical User Interface (GUI) of \acrshort{design}.}
    \label{fig:GUI}
\end{figure}

We have provided a Graphical User Interface (GUI) to facilitate the study of near-field channel modeling for the next generation of communication systems. Fig. \ref{fig:GUI} illustrates the GUI main window. We have made the core codebase of the \acrshort{design}
simulator readily accessible in \href{https://github.com/vahidyazdnian1378/NirvaWave}{GitHub repository}.

\subsection{Input Parameters}
In \acrshort{design}, users can specify various environmental factors, including dimension, resolution, blockages, reflectors, and RISs, as well as the properties of the transmitter (TX) and receiver (RX) antenna arrays, and their phase configurations. We have included a complete instruction of input parameters in the Github repository. The spatial resolution is custom-defined through \textit{Dimension} and \textit{Resolution} inputs. We note that a minimum spatial resolution of $\lambda/2$ is required to correctly capture EM propagation. Users can also define reflectors and blockages in the environment by specifying their center location, length, orientation, thickness, and reflection coefficient. For rough surfaces, the statistical roughness parameters $h_{rms}$ and $L_c$ can be set. Further, users can configure an RIS into the environment at any location with the desired phase profile.

%To incorporate an RIS into the environment, users can specify the reflector in the form of an RIS and import the corresponding phase shifts for each element via a text file. The number of phase configurations specified in the text file must be at least equal to the number of discrete points on the RIS plane, as determined by the resolution. It is advisable to define the phase configurations according to the RIS length and environment resolution. Additionally, suppose that the number of intended elements on the RIS is fewer than those specified by the resolution. In that case, the actual phase shifts should be repeated for the RIS elements in the simulation environment to modify the received E-fields consistently across each element.

The GUI allows for configuring the TX antenna properties and the operating frequency. \acrshort{design} imposes no restrictions on frequency or dimension, as it is accurate in both Near-Field and Far-Field regimes. We emphasize that unlike EM simulators like HFSS and CST that allow for custom antenna geometry and design, \acrshort{design} is not an antenna simulator. Specifically, we assume ideal isotropic antennas at the transmitter aperture that can adopt any custom phase or amplitude profile by importing a text file, the details of which can be found on the GitHub repository. Additionally, \acrshort{design} allows for simulations of arbitrary near-field wavefronts, including Bessel beams and Airy beams. Users can also input their custom complex E-field distribution on TX aperture and simulate their behavior in free-space and inhomogeneous mediums. Finally, in  \acrshort{design} users can define multiple RX antenna array properties, including center location, length, orientation, and the number of elements, to calculate the power at the intended UE locations. By default, the RX antenna is assumed to be fully digital. For analog arrays, users should import a text file containing the weight vector configuration.

\subsection{Output Files}
The \acrshort{design} provides an accurate yet efficient wave propagation profile based on the underlying physics principles outlined before. The output involves coverage maps and received power at specified User Equipment (UE) locations. Additionally, the calculated electric field at each location can be saved for further analysis. Examples of radiation maps are provided in Sec.~\ref{sec:evaluation}.

\begin{figure}[t]
    \centering
    \begin{subfigure}[b]{0.24\textwidth}
        \centering
        \includegraphics[width=0.7\textwidth]{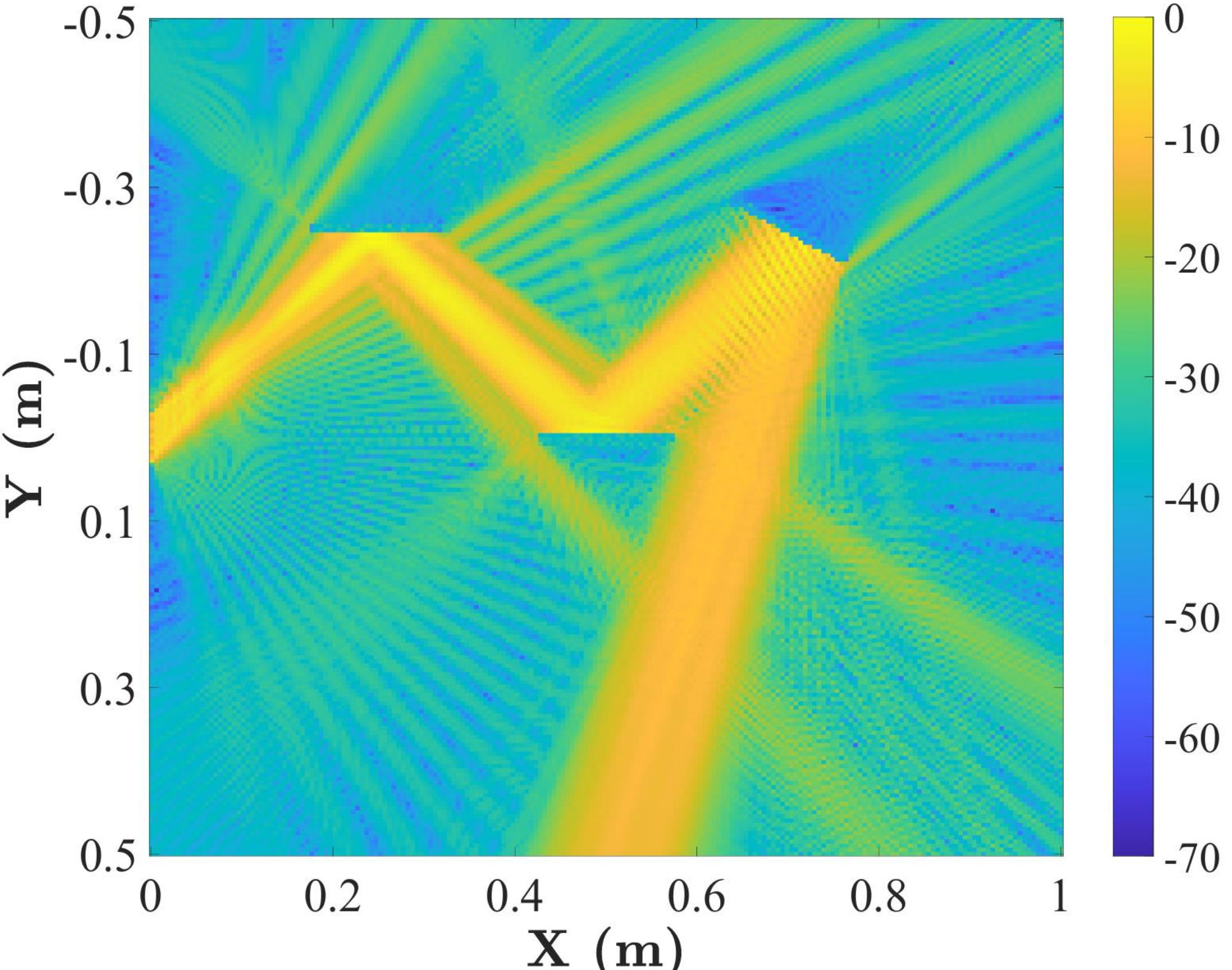}
        \caption{Feko coverage map}
        \label{fig:feko_3ref}
    \end{subfigure}
    \hfill
    \begin{subfigure}[b]{0.235\textwidth}
        \centering
        \includegraphics[width=0.7\textwidth]{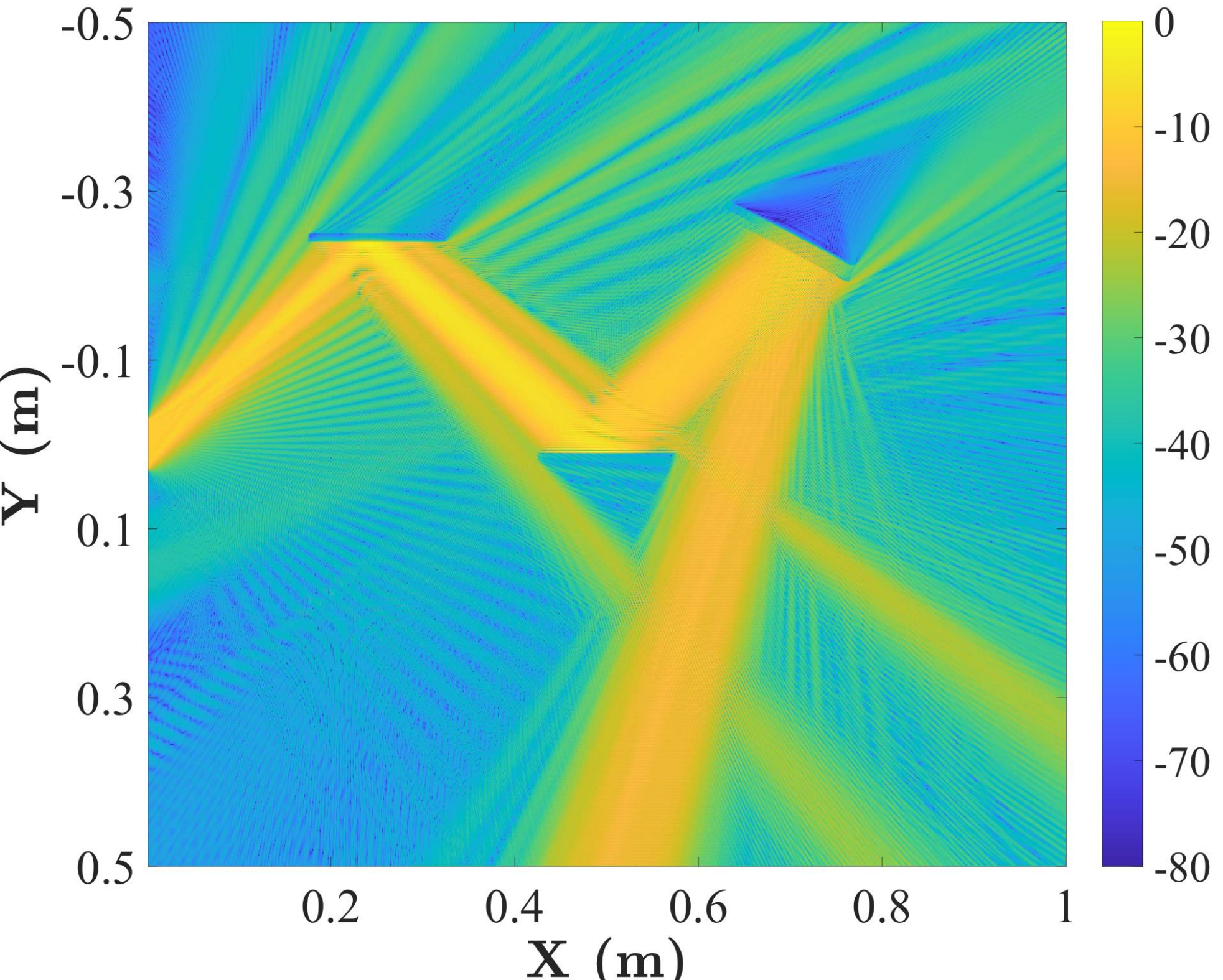}
        \caption{\acrshort{design} coverage map}
        \label{fig:design_3ref}
    \end{subfigure}
    \vspace{-1mm}
    \caption{Coverage map results from \acrshort{design} and Feko simulations in an environment with three reflectors.}
    \label{fig:3ref}
    \vspace{-6mm}
\end{figure}

\subsection{Large Data Dataset Collection}
The fast simulation run-time, the ultimate flexibility in transmitted electric fields, and on-demand wireless mediums make \acrshort{design} a promising tool for data collection at scale. Such massive data collection plays a pivotal role in deepening our understanding of near-field wireless channels in this untapped territory, as well as developing new data-driven and AI-assisted models that can estimate or predict how EM waves evolve in complex environments. To assist users in getting started, we have included a basic script file in GitHub that serves as a guide. This script can be easily customized to meet the unique requirements of different applications.

%% file: evaluation.tex
\section{Simulator Evaluation}\label{sec:evaluation}
In this section, we evaluate the performance of the \acrshort{design} simulator in various environmental configurations and wavefront designs. We compare our result with  Altair Feko\cite{feko}, a sophisticated EM simulation software. It should be noted that the simulation environment in Feko is 3D with 2D antenna apertures. However, we only solve the EM wave propagation in the plane of interest to compare the output results to \acrshort{design}. We run all the simulations at 100 GHz.

\textbf{Metric.} We exploit two complementary metrics for evaluation purposes. The first metric is Root Mean Squared Error (RMSE), which measures the point-wise difference between the normalized E-field intensities from Feko and \acrshort{design}, ranging from 0 (no difference) to 1 (maximum difference). The second metric aims to evaluate the structural similarity between the two E-field heatmaps. To this end, we can calculate the 2D cross-correlation of normalized $N\times N$ matrices $|\textbf{E}_{Feko}|$ and $|\textbf{E}_{sim}|$, i.e., $\mathcal{\textbf{R}}= |\textbf{E}_{Feko}| \star |\textbf{E}_{sim}|$, where $\mathcal{\textbf{R}}$ is a $(2N-1) \times (2N-1)$. We use the maximum value in $\mathcal{\textbf{R}}$ as the metric of similarity between coverage maps which ranges from -1 to 1, with 1 showing a perfect match. Hence, a cross-correlation close to 1 demonstrates a higher structural similarity, which RMSE alone cannot capture.

\textbf{Simulation Run-Time.} EM simulation software like Feko that models EM wave propagation by solving Maxwell's equations is accurate, albeit extremely time-consuming in practice. The run-time issue is exacerbated as the number of reflectors or the TX aperture size increases. In a basic environment with two reflectors and a 10 cm TX aperture, the simulation in \acrshort{design} is over 40 times faster. More importantly, while the runtime of large EM simulators increases exponentially with the TX aperture size, the runtime in \acrshort{design} remains almost unaffected. For instance, with a 20 cm TX aperture and one reflector, \acrshort{design}'s run time is $200+$ times faster than Feko. In more complex environments, the efficiency of \acrshort{design} compared to EM simulators becomes even more pronounced.

\subsection{Performance of \acrshort{design} in Reflector Simulations}
First, we evaluate the performance of \acrshort{design} in capturing the near-field channels under the presence of reflectors. We systematically try 40 different environmental configurations such that the reflector is in the near-field of the TX in all cases. We also tested different roughness parameters to assess the performance of \acrshort{design} under diffuse rough scattering.  

\subsubsection{Specular Reflection} 
\begin{figure}[t]
    \centering
    \begin{subfigure}[b]{0.24\textwidth}
        \centering
        \includegraphics[width=0.7\textwidth]{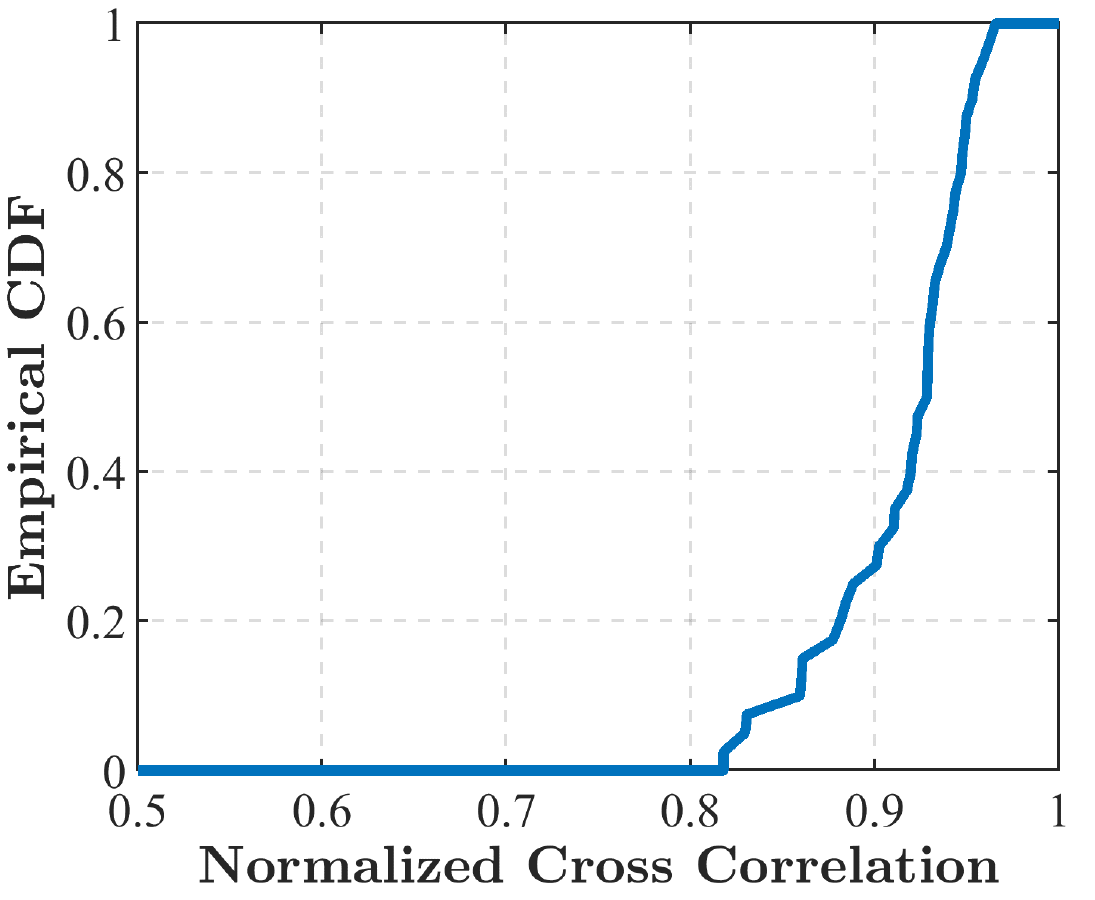}
        \caption{Normalized correlation }
        \label{fig:norm_corr_ref}
    \end{subfigure}
    \hfill
    \begin{subfigure}[b]{0.24\textwidth}
        \centering
        \includegraphics[width=0.7\textwidth]{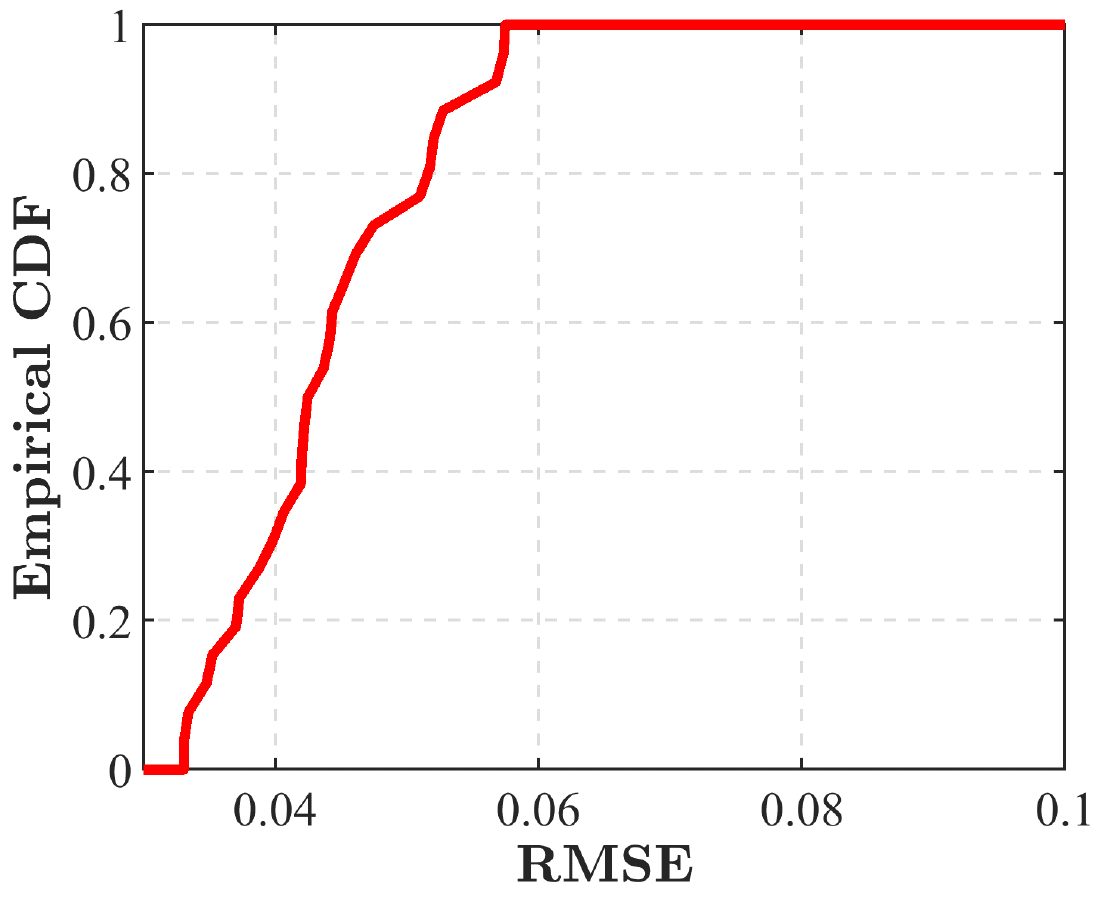}
        \caption{RMSE}
        \label{fig:rmse_ref}
    \end{subfigure}
    \caption{Evaluation of \acrshort{design} accuracy in near-field channel modeling in comparison with Feko in different environments.}
    \vspace{-4mm}
    \label{fig:ref_evaluation}
\end{figure}

We simulated 18 scenarios with a single reflector, 14 cases with two reflectors, 3 scenarios with three reflectors, 3 environments with a single blocker, and two special cases with non-Gaussian transmitted wavefronts (Airy Wavefront and Bessel Beam). Fig.~\ref{fig:3ref} illustrates the simulation results from \acrshort{design} and Feko for an environment with three reflectors. We observe that visually the heatmaps match pretty well. Additionally, Fig.~\ref{fig:ref_evaluation} presents the empirical CDFs of RMSE and normalized correlation for all 40 configurations. RMSE is under 0.06 for all settings, with 80\% achieving below 0.05. Normalized cross-correlation is above 0.8 across all configurations, with 80\% exceeding 0.9. The results demonstrate \acrshort{design}'s ability to accurately simulate reflection and blockage for any arbitrary primary electric field distributions.

\begin{figure}[t]
    \centering
    \includegraphics[width=0.35\textwidth]{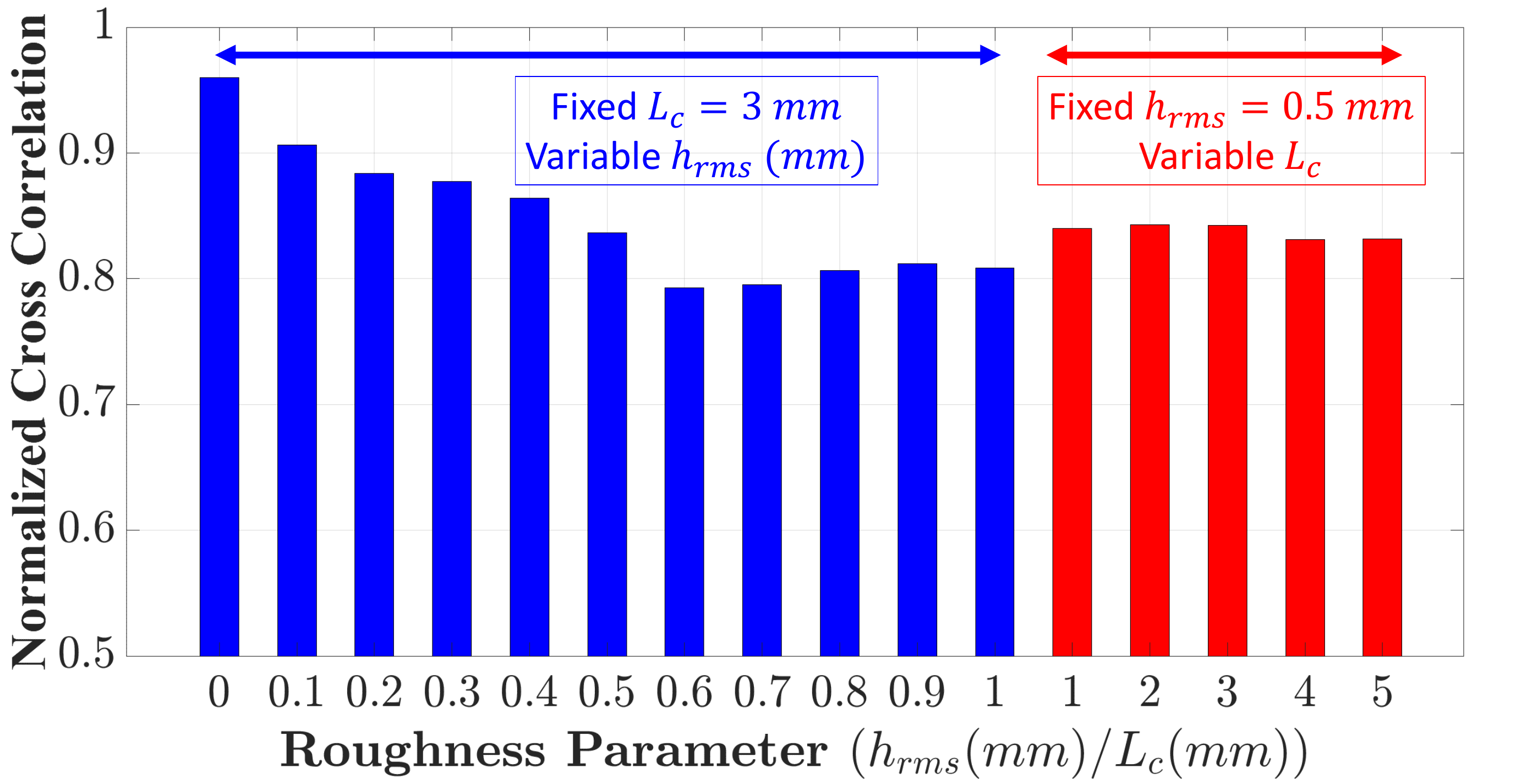}
    \vspace{-2mm}
    \caption{Normalized cross correlation between \acrshort{design} and Feko in diffuse scattering simulations. Blue bars represent fixed $L_c = 3$ mm and variable $h_{rms} \, (mm)$ while red bars represents fixed $h_{rms} = 0.5$ mm and variable $L_c \, (mm)$.}
    \label{fig:rough_bar}
    \vspace{-6mm}
\end{figure}

\begin{figure*}[t]
    \centering

    % First subfigure
    \begin{subfigure}[b]{0.32\textwidth}
        \centering
        \includegraphics[width=0.49\textwidth]{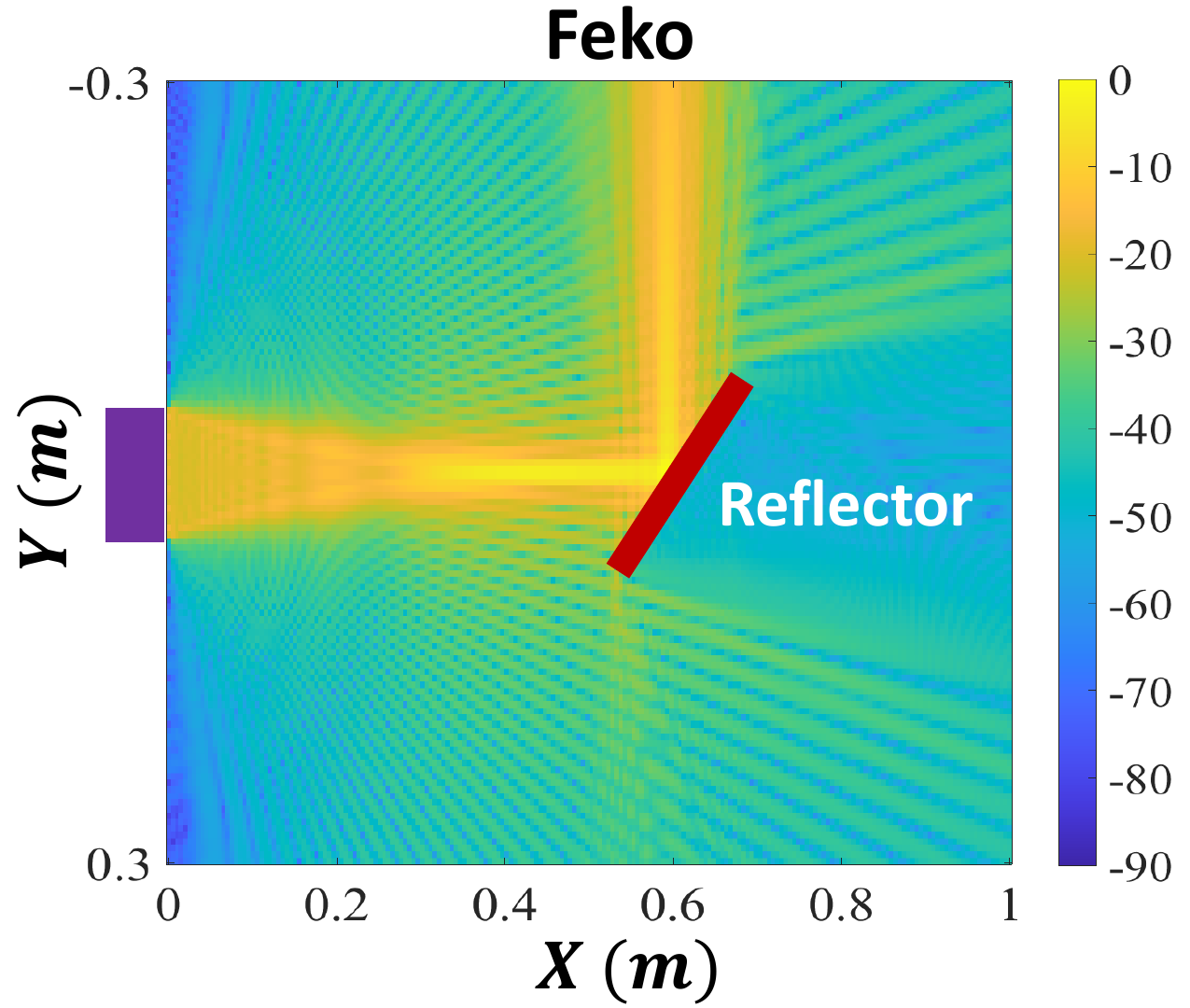}
        \includegraphics[width=0.49\textwidth]{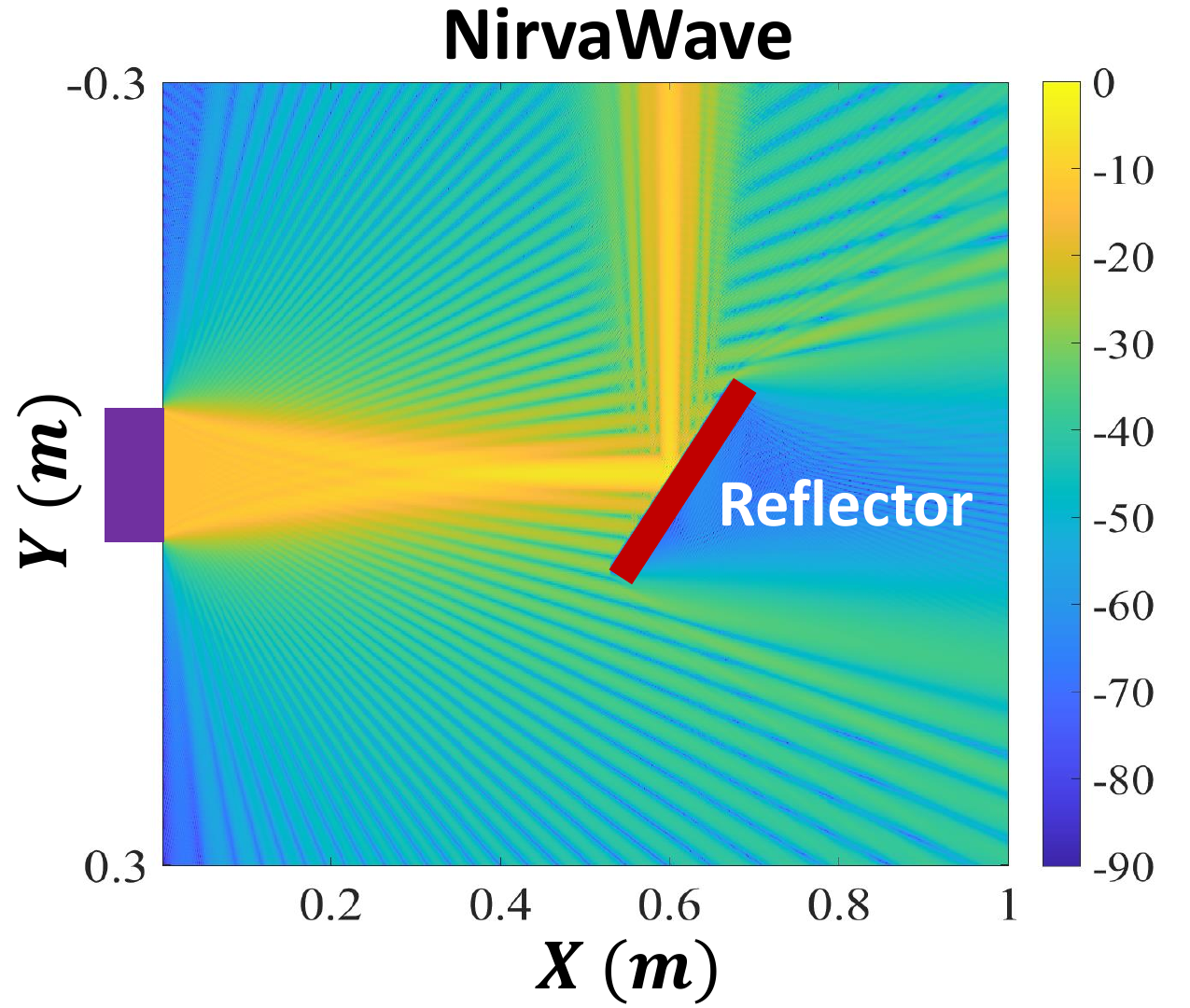}
        \vspace{-6mm}
        \caption{Focused Beam}
        \label{fig:Focused_Beam}
    \end{subfigure}
    \hfill
    \begin{subfigure}[b]{0.32\textwidth}
        \centering
        \includegraphics[width=0.48\textwidth]{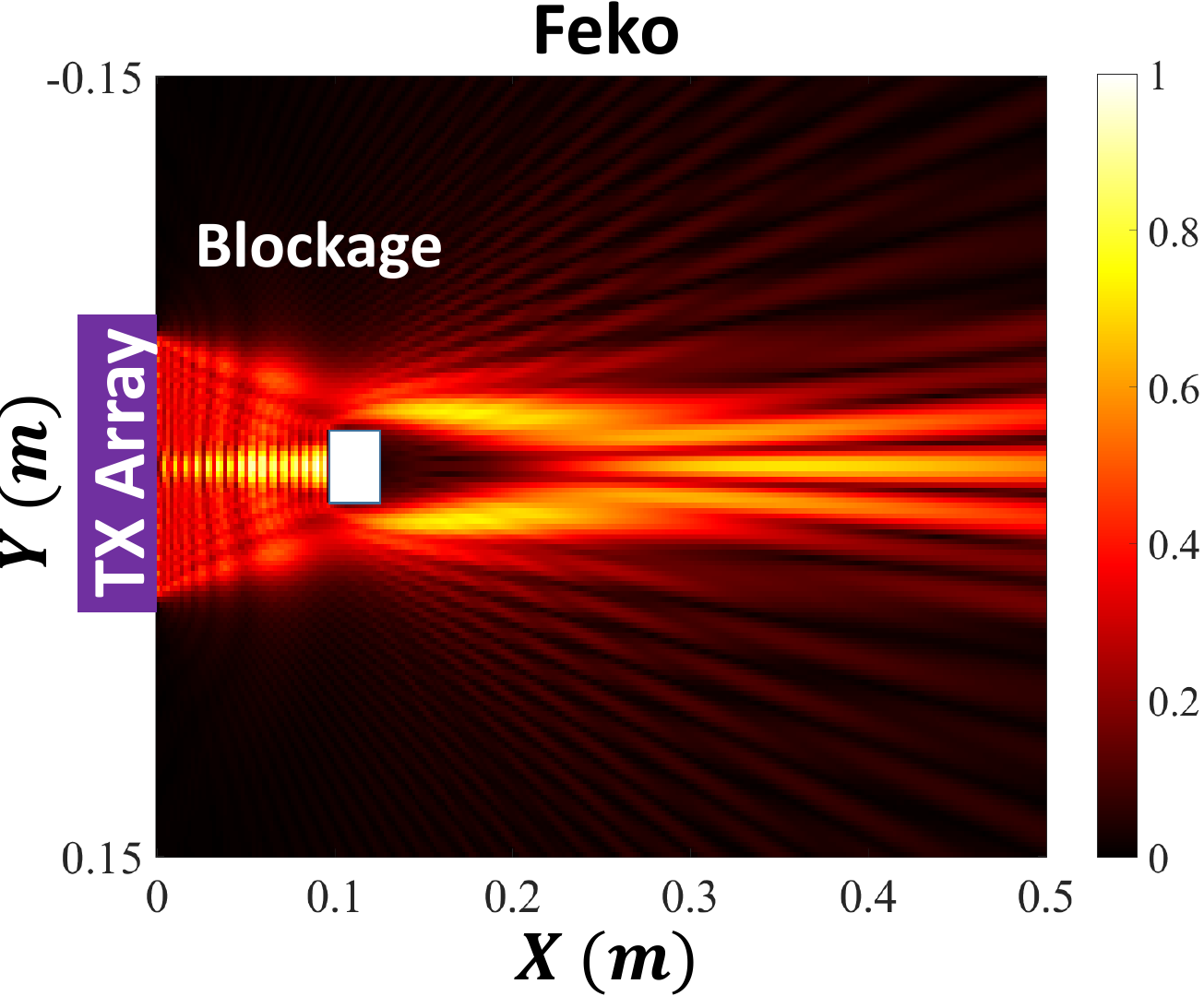}
        \includegraphics[width=0.49\textwidth]{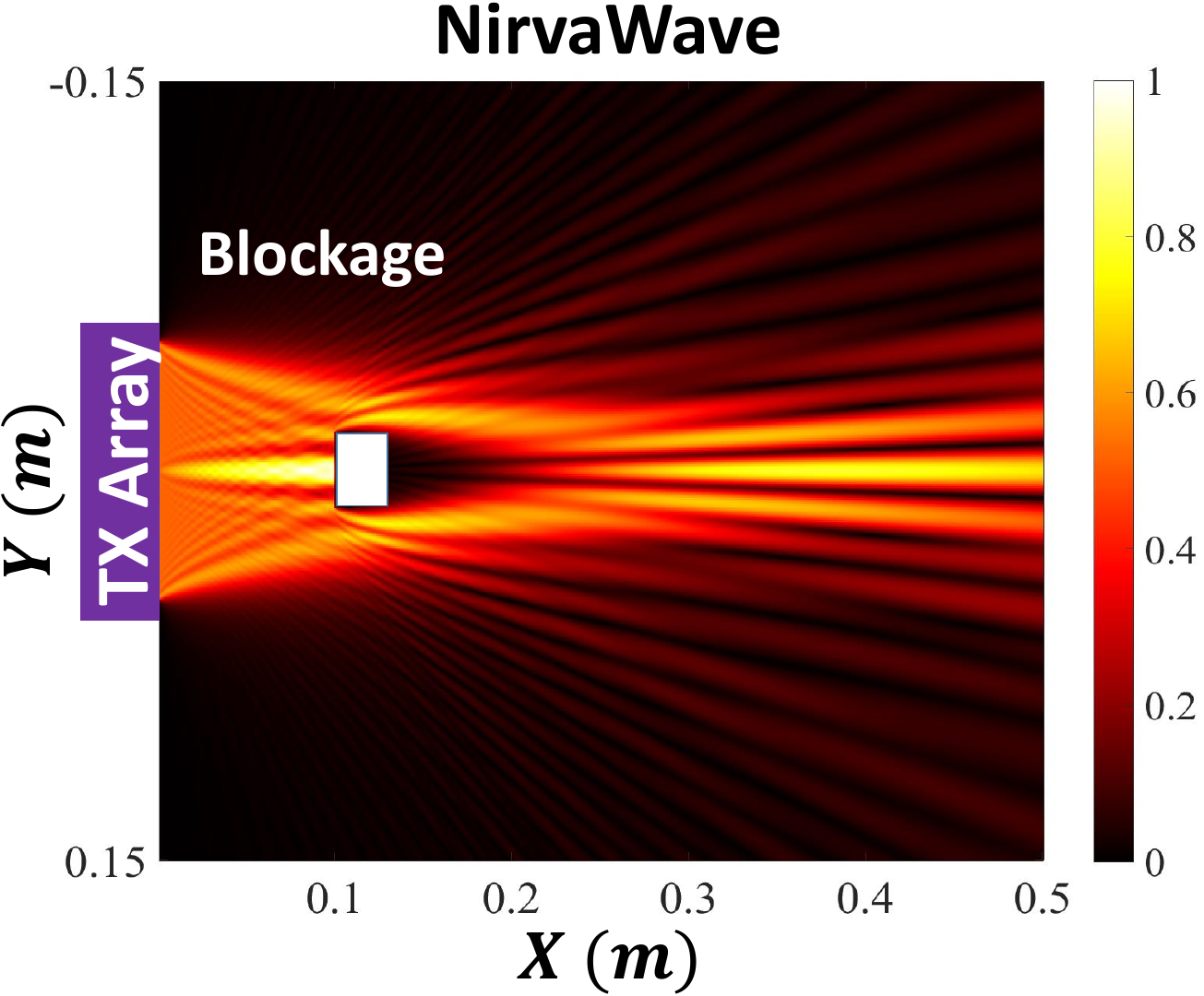}
        \vspace{-6mm}
        \caption{Bessel Beam}
        \label{fig:Bessel_Beam}
    \end{subfigure}
        \hfill
    \begin{subfigure}[b]{0.32\textwidth}
        \centering
        \includegraphics[width=0.49\textwidth]{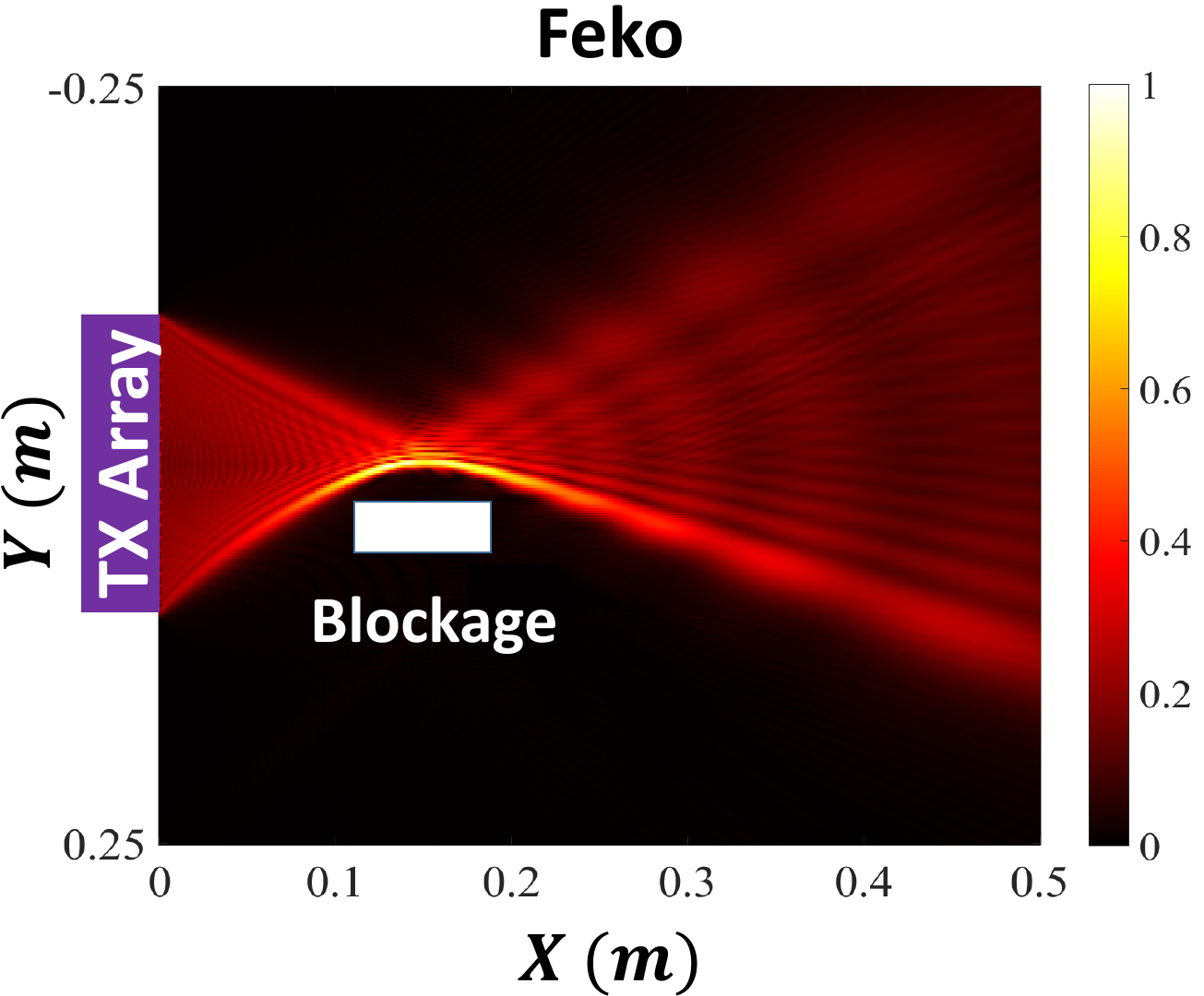}
        \includegraphics[width=0.49\textwidth]{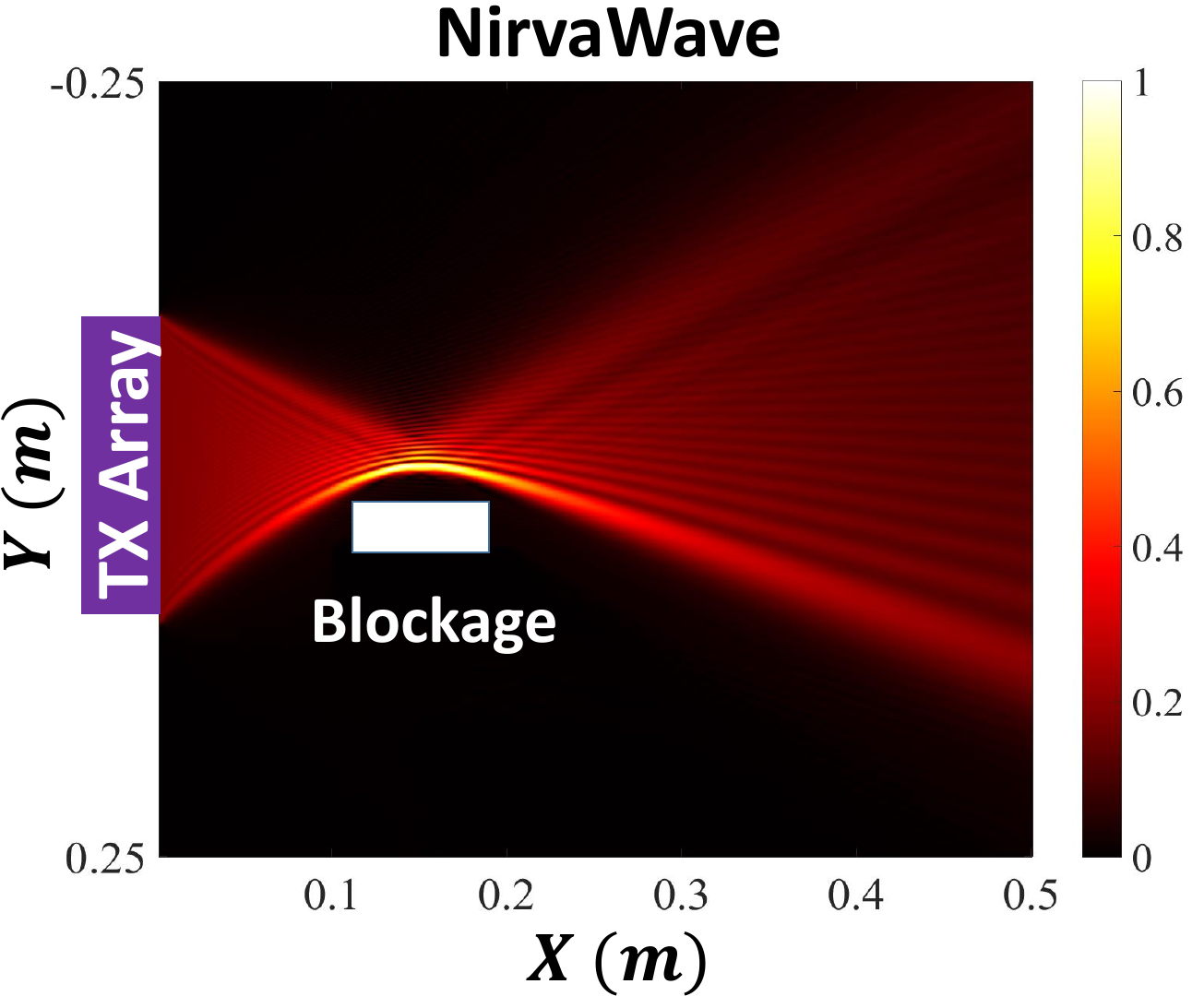}
        \vspace{-6mm}
        \caption{Airy Beam}
        \label{fig:Airy_Beam}
    \end{subfigure}
    \vspace{-2mm}
    \caption{Emerging near-field wavefornts simulation results in Feko and \acrshort{design}.}
    \label{fig:main_figure}
\end{figure*}
\vspace{-3mm}
\subsubsection{Diffuse Rough Scattering}
To simulate diffuse rough scattering, we generate a random height distribution, representing the surface height perturbations, based on a specific pair of $h_{rms}$ and $L_c$ values and import the rough surface into Feko. For simulating the same surface in \acrshort{design}, we sample the generated height distribution and convert it into corresponding phase shifts on the reflector plane. We conducted 15 simulations with varying roughness parameters: 10 with $L_c = 3$ mm and varying $h_{rms}$ from 0 to 1 mm, and 5 with $h_{rms} = 0.5$ mm and varying $L_c$ from 1 to 5 mm. Fig.~\ref{fig:rough_bar} presents the normalized cross-correlation between the simulations in Feko and \acrshort{design} for all 15 scenarios. We observe that the normalized cross-correlation exceeds $0.80$ in all settings, with an average value of $0.84$. These findings demonstrate \acrshort{design}'s capability to accurately capture the diffuse rough scattering phenomenon in the near-field region of the THz Tx antenna arrays. However, as surface roughness increases, the discrepancies between the two simulators slightly grow, mainly due to the increased impact of random diffuse scattering in 3D for rougher surfaces. Nevertheless, for most natural indoor/outdoor surfaces, $h_{rms}<0.7$~\cite{shen2023scattering}.
\vspace{-0.5mm}
\subsection{Unique Near-field Wavefronts}
One of the unique advantages of \acrshort{design} is the ability to simulate arbitrary transmit wavefronts, unlike state-of-the-art simulators that solely consider conventional Gaussian beams.
Here, we simulate three special near-field wavefronts, namely focused beam, Airy beam, and Bessel beam, to evaluate the performance of \acrshort{design} in capturing the special characteristics of these wavefronts in inhomogeneous mediums:

\textit{(i)} First, we implement a focused beam in the presence of a near-field reflector. To this end, we use a 10 cm TX array at 100 GHz yielding a near-field distance of around 3 meters. We place a reflector at $(x, y) = (0.6 \text{ m}, 0 \text{ m})$ with $R_L = 0.2$ m and $\theta_r = 45^\circ$, i.e., the reflector is in the near-field of the array. We configure the focused beam to have a focal length of $0.6$ m. The results from \acrshort{design} and Feko in this environment are shown in Fig. \ref{fig:Focused_Beam}. We observe that \acrshort{design} matches well with EM software Feko. \textit{(ii)} The Bessel beam is emerging due it its non-diffraction properties (in a certain regime) and its self-healing property, which allows the beam to reconstruct even after being blocked by an obstacle~\cite{singh2022bessel}. To simulate Bessel beams, we use a TX aperture dimension of 10 cm at 100 GHz and adopt the phase configuration corresponding to a Bessel beam non-diffraction angle\cite{singh2022bessel} of $\alpha = 5^\circ$, with a blockage placed at $(x, y) = (0.1 \text{ m}, 0 \text{ m})$. The results from \acrshort{design} and Feko in this environment are illustrated in Fig. \ref{fig:Bessel_Beam}. We observe that the wavefront before and after the obstruction is captured accurately by \acrshort{design}. \textit{(iii)} Finally, we implemented an Airy beam that has a curved trajectory in space making it an interesting solution to curve around obstacles in the environment and mitigate blockages. We simulate an aperture of size 20 cm at 100 GHz with the Airy phase function~\cite{latychevskaia2016creating} to achieve a curvature parameter of $\beta = -4.5$ and a focal length of $f = 0.15$ m. We place a blockage at $(x, y) = (0.15 \text{ m}, -0.07 \text{ m})$ with a thickness of 5 cm and a length of $R_L = 0.1$ m. Fig.~\ref{fig:Airy_Beam} demonstrates that \acrshort{design} captures the correct trajectory of near-field waves as they curve around the obstacle.

These results demonstrate \acrshort{design}'s ability to capture the near-field characteristics of important and arbitrary wavefronts in the presence of blockages and reflectors in the wireless setting, which is essential for studying the sub-THz near-field channel for the next generation of communication systems.

%% file: conclusion.tex
\section{Conclusion}
This paper introduces \acrshort{design}, an open-source simulator developed for modeling near-field EM wave propagation based on scalar diffraction theory. \acrshort{design} offers a precise and efficient framework for analyzing the near-field propagation and interaction of EM waves in various user-defined wireless mediums incorporating blockers, reflectors, and scatterers. It allows users to set TX antenna array properties and specify the corresponding complex E-field distribution either through arbitrary user-defined inputs or by selecting from a range of implemented emerging near-field wavefronts. With its user-friendly interface, \acrshort{design} facilitates the study of mmWave and sub-THz near-field channels while ensuring efficiency, scalability, and accuracy, paving the way for the development of model-driven and data-driven techniques to address the challenges of wireless communications in these high-frequency regimes.

% This paper presents NirvaWave, an open-source simulator for modeling near-field EM wave propagation using scalar diffraction theory. NirvaWave provides an efficient and accurate framework for analyzing near-field wave interactions in various user-defined wireless mediums. It allows users to configure TX antenna array properties and specify complex E-field distributions, either through user-defined inputs or by choosing from implemented near-field wavefronts. The simulator can model near-field channels in diverse environments, including scenarios with blockers, reflectors, and RISs. With its user-friendly interface, NirvaWave supports the study of mmWave and sub-THz channels, facilitating the development of techniques to overcome challenges in high-frequency wireless communications.